\documentclass[doc,apacite]{apa6}\usepackage[]{graphicx}\usepackage[]{color}
\pdfoutput=1 
\makeatletter
\def\maxwidth{ %
  \ifdim\Gin@nat@width>\linewidth
    \linewidth
  \else
    \Gin@nat@width
  \fi
}
\makeatother

\definecolor{fgcolor}{rgb}{0.345, 0.345, 0.345}

\usepackage{framed}
\makeatletter
 {\par\unskip\endMakeFramed%
 \at@end@of@kframe}
\makeatother

\definecolor{shadecolor}{rgb}{.97, .97, .97}
\definecolor{messagecolor}{rgb}{0, 0, 0}
\definecolor{warningcolor}{rgb}{1, 0, 1}
\definecolor{errorcolor}{rgb}{1, 0, 0}
\newenvironment{knitrout}{}{} % an empty environment to be redefined in TeX

\usepackage{alltt}
\usepackage[american]{babel}
\usepackage[utf8]{inputenc}
\usepackage[T1]{fontenc}
\usepackage[kerning=true]{microtype}
\usepackage{csquotes}
\usepackage{paralist}
\usepackage{longtable}
\usepackage{color}

\usepackage{apacite}

\title{False Positives and Other Statistical Errors in Standard Analyses of Eye Movements in Reading}
\shorttitle{False positives in reading experiments}
\leftheader{von der Malsburg, Angele}

\note{Revision 2 -- October 10 2016}

\twoauthors{Titus von der Malsburg}{Bernhard Angele}
\twoaffiliations{Department of Psychology and Department of Linguistics, UC San Diego}
                {Department of Psychology, Bournemouth University}

\authornote{We dedicate this article to the memory of Keith Rayner who commented on an early version of this work.  We also thank Shravan Vasishth, the members of the Rayner Lab, and audience members at the CUNY, AMLaP, and ECEM conferences for insightful comments.  The code and simulation results will be made available online upon publication of this article and the URL will be inserted here.  This research was conducted using free software published under the GNU General Public License, particularly the R system for statistical computing, the Slurm Workload Manager, GNU Emacs, the TeX Live typesetting system, the knitr package for literate programming, and GNU/Linux.  Titus von der Malsburg was funded by a Feodor Lynen Research Fellowship awarded by the Alexander von Humboldt Foundation and by NIH grant HD065829 awarded to Roger Levy and Keith Rayner.  Please direct correspondence to Titus von der Malsburg (malsburg@ucsd.edu).}


\abstract{In research on eye movements in reading, it is common to analyze a number of canonical dependent measures to study how the effects of a manipulation unfold over time.  Although this gives rise to the well-known multiple comparisons problem, i.e.~an inflated probability that the null hypothesis is incorrectly rejected (Type I error), it is accepted standard practice not to apply any correction procedures.  Instead, there appears to be a widespread belief that corrections are not necessary because the increase in false positives is too small to matter.  To our knowledge, no formal argument has ever been presented to justify this assumption.  Here, we report a computational investigation of this issue using Monte Carlo simulations.  Our results show that, contrary to conventional wisdom, false positives are increased to unacceptable levels when no corrections are applied.  Our simulations also show that counter-measures like the Bonferroni correction keep false positives in check while reducing statistical power only moderately.  Hence, there is little reason why such corrections should not be made a standard requirement.  Further, we discuss three statistical illusions that can arise when statistical power is low, and we show how power can be improved to prevent these illusions.  In sum, our work renders a detailed picture of the various types of statistical errors than can occur in studies of reading behavior and we provide concrete guidance about how these errors can be avoided.}
\IfFileExists{upquote.sty}{\usepackage{upquote}}{}
\begin{document}

\maketitle

A key advantage of using eye-tracking methods in reading research is the wealth of the data that can be collected. The entire sequence of fixations that a participant produces during a trial is recorded and, in theory, available for analysis. In practice, some standard aggregate measures have been established to adequately summarize this wealth of data \cite{Rayner1998}. Most commonly, the analysis region for which these measures are computed will contain one word or phrase within a longer sentence. Over the last decades, many different measures have been proposed, but there are a handful of standard measures that almost every eye-tracking study reports. In order to compute these measures, the fixation sequence is divided into the first pass, consisting of the fixations occurring when a reader first enters a word from the left, and the second pass, consisting of the fixations occurring when a word is revisited \cite<see>[for a discussion of the distinction between early and late measures]{VasishthEtAl2013}. First pass and second pass fixations are then filtered and combined to form the standard aggregate measures:

\begin{APAenumerate}
\item
  \emph{First fixation duration (FFD)}: the duration of the very first fixation a reader made on a region, regardless of whether it was refixated thereafter or not.
\item
  \emph{Single fixation duration (SFD)}: the same as first fixation duration, but only counting cases where a region was not refixated during first pass.
\item
  \emph{Gaze duration (GZD)}: the sum of the durations of the first fixation and all refixations during first pass, that is, before the gaze left the word for the first time.
\item
  \emph{Go-past duration (GPD)}: also known as regression path duration, the sum of all fixations from when a reader's gaze first entered a word from the left, including all refixations and all regressive fixations to prior words, until it left the word towards the right.
\item
  \emph{Total viewing time (TVT)}: the sum of the durations of all fixation on the word, regardless of whether they occurred during the first or second pass.
\end{APAenumerate}

Again, the above list of measures is not exhaustive, but the five measures described are those most commonly used. From the description of these measures it should already be apparent that they are all correlated to varying degrees. These correlations, combined with the fact that several of these measures are potentially informative, lead to a problematic situation.

Let's assume that a researcher wants to test whether a certain experimental manipulation has an impact on reading behavior. They will then compute the condition means for each measure and, in order to test whether those means are significantly different, compute a series of ANOVAs or linear mixed-effects models, one for each measure.  In the absence of a specific hypothesis about the impact of the manipulation on the reading process, the simplest and most commonly used decision strategy is to conclude that reading is affected by the manipulation if at least one of these analyses indicates a significant difference in means.  Which measure is affected is not important under this decision strategy; at best, it might provide some additional information about the time course of the affected process.

The problem with this decision strategy is the following: Assuming that significance is determined using an \emph{alpha} threshold of $0.05$, each test produces a positive result with a probability of 5\% even if there was no true effect.  Assuming further that four statistically independent eye-tracking measures are tested, the probability that at least one of these tests produces a false positive result increases to $1 - 0.95^4 = 0.185$.  This means that the probability of finding a false positive result would be 18.5\% instead of the conventionally accepted 5\%.

Of course, in reality, eye-tracking measures are not quite independent as the different fixation time measures are typically highly correlated.  For example, the correlation between first fixation duration (FFD) and gaze duration (GZD) is determined by the rate at which readers refixate. If readers do not refixate at all, FFD and GZD are identical, in other words, their correlation is 1.  The more readers refixate, the lower the correlation between FFD and GZD becomes. However, since the fixations that are used to compute FFD form a subset of those used to compute GZD, the correlation will typically not reach zero.

The fact that the canonical eye-tracking measures are correlated implies that the probability of a falsely declared effect is not quite as high as the 18.5\% calculated above.  We suspect that this is one reason why many reading researchers draw conclusions as if the false positive rate had only been 5\%.  However, whether this simplifying assumption is warranted or not is unclear.  If the correlations between fixation time measures were always 1, multiple comparisons would indeed not be cause for concern; all tests would necessarily produce the same result such that no test beyond the first could possibly produce additional false positives.  However, in that case, analyzing different fixation time measures would be completely redundant.  Eye movement researchers are clearly aware that different fixation time measures share some information, but that each measure also contributes unique information.  Thus, the true false positive rate in an eye-tracking study with four dependent measures lies somewhere between 5\% and 18.5\%.

If the rate of false positives is inflated to unacceptable levels due to multiple comparisons, adjustments to the decision making process are needed.  The text-book solution for that is the Bonferroni correction \cite{Bonferroni1936}.  All a researcher has to do in order to apply this correction is to divide the $\alpha$ threshold for determining significance (typically 0.05) by the number of tests that were performed and to use the resulting number as the new threshold.  In the present scenario, that means that significance would be determined using the threshold $\alpha = 0.05/4 = 0.0125$.  If the threshold is lowered in this way, the probability of finding at least one false positive result in any of the multiple statistical tests is 5\%.  We then say that the family-wise error rate is controlled, the family being the set of tested hypotheses: there is an effect in FFD, there is an effect in GZD, and so on.

Unfortunately, the Bonferroni correction is not entirely appropriate for analyses of eye-tracking measures because it assumes that the statistical tests are independent, which, as we discussed above, they are typically not.  The Bonferroni correction may therefore be too conservative, which means that the false positive rate will be even lower than 5\%.  While this is not in itself a problem, any reduction of false positives also comes at the price of reduced statistical power, i.e. a reduced ability to detect true effects.  Hence, it is desirable to reduce false positives to the conventional 5\% but not more because we would otherwise sacrifice more statistical power than necessary in order to control the family-wise error rate.

Concerns about loss of statistical power may be another reason for why reading researchers do not apply corrections for multiple comparisons such as the Bonferroni correction.  However, this leads us to a dilemma: applying a Bonferroni correction may lead to an unnecessary loss of statistical power (Type II errors), but the alternative, applying no correction at all, will inflate the probability of false rejections of the null hypothesis (Type I errors).  The question is: on which side should we err?

\citeNP{SimmonsEtAl2011} argue that false rejections of the null hypothesis (Type I errors) are particularly costly and have the potential to threaten the integrity of the scientific discourse much more than false negative results (Type II errors).  \citeNP{McElreathSmaldino2015} illustrate and substantiate this point using a formal model of the population dynamics of scientific discovery.  Their results show that false positives can easily proliferate and clog up the scientific discourse because the resources for weeding them out are spread too thinly.  However, whether more harm is done by over-correcting or by not correcting at all crucially depends on the true rate of false positives and on how much power is really sacrificed when corrections are applied.  In the following, we present the results of a Monte Carlo simulation that sheds light on these questions.

\subsection{Overview}
\label{overview}

The purpose of the simulations presented below was two-fold: First, we determined by how much the rate of false positives is elevated when multiple statistically dependent eye-tracking measures are tested without correcting for multiple comparisons.  Second, we investigated by how much statistical power is reduced through the application of two correction methods.

The first alternative decision strategy was to lower the alpha threshold using the Bonferroni method and to declare an effect of the manipulation only if at least one of the eye-tracking measures showed an significant difference between conditions with a \emph{p}-value at or below that lowered level.\footnote{We also tested the Holm-Bonferroni correction, an alternative to the Bonferroni correction that preserves slightly more power, but the difference between these methods was so small that we decided to focus on the simpler and more well-known Bonferroni correction.}

The second alternative decision strategy was of a more information nature and we are therefore going to call it the rule-of-thumb criterion: In order to declare an effect reliable, this criterion requires that at least two eye-tracking measures show an effect with \emph{p}-values at or below $0.05$.  To our knowledge, this criterion has not explicitly been used in the published literature.  However, it may be tempting to use this criterion in more informal settings – perhaps even unconsciously – due to its simplicity and because it intuitively makes sense to think that an effect is more reliable when it emerges two measures instead of just one.  While this criterion is stricter than our baseline criterion (just one measure with $p\leq0.05$), we will see that it performs poorly at best and that it may even fail catastrophically, namely in the situation where the effect can only emerge in one measure.

To assess false positive rates and statistical power under the baseline strategy (no correction) and the two alternative strategies (Bonferroni correction and rule-of-thumb criterion), we employed a Monte Carlo simulation approach. The basic idea is to generate many artificial data sets with properties resembling those found in real eye-tracking data and to analyze these data sets in order to determine false positive rates and statistical power. This is possible because we know the population-level effect present in the artificial data and thus can ascertain whether the outcome of a statistical test is accurate or a false positive or a false negative.

During each iteration of the simulation, we generated a number of artificial data sets that contained the eye movement measures mentioned above (FFD, GZD, GPD, TVT; we did not use SFD since it is just a subset of FFD) for a number of artificial subjects and items. One of these data sets had no effect of the hypothetical manipulation and the other data sets had increasingly larger effects (2.5, 5, 10, 20, 40, 80\,ms). This allowed us to assess the rate of false positives, i.e. the proportion of simulation runs in which no true effect was present, yet the overall null hypothesis -- that there was no effect in any of the eye-tracking measures -- was rejected.  It also allowed us to assess statistical power, i.e. the proportion of simulation runs in which true effects were present and the overall null hypothesis was correctly rejected.  By repeating these simulations many times, we could calculate the expected rates of false positives and statistical power with high precision and study how both depend on various parameters such as the numbers of subjects and items.

In each iteration, the presence of an effect was determined using all three decision criteria discussed above. The \emph{p}-values were computed by fitting linear mixed models and conducting likelihood-ratio tests comparing models with the factor for condition vs. models without this factor \cite<see>[for a similar approach]{BarrEtAl2013}.

\section{Materials and methods}

Our approach to estimating false positive rates and power requires that we generate a large number of artificial data sets. For the results to be valid, it is important that these data sets have statistical properties similar to those found in real eye-tracking data. Given the complex nature of eye movements in reading, this is not entirely trivial. After all, the ability to generate perfectly realistic eye movement data presupposes complete knowledge about the principles and mechanisms underlying the reading process. Hence, we have to make a number of simplifying assumptions: \begin{seriate}
  \item The target word is fixated at least once.
  \item The target word can be refixated during first pass leading to
    gaze durations being longer than first fixation durations.
  \item A regressive eye movement can occur during the first pass
    leading to elevated go-past durations.
  \item A word can be refixated after the first pass leading to
    increased total viewing times.
  \item The artificial subjects differ in the average duration of
    their first fixations.
  \item The artificial items differ in the average duration of the
    first fixations they elicit.
\end{seriate}

A lot of characteristics of eye movements in reading are not captured by this model.  However, as we will see below, the eye-tracking measures generated with this model are sufficiently similar to real data.  In particular, the correlations between the various eye-tracking measures are largely preserved. This is important because, as discussed above, these correlations influence false positive rates.

\subsection{Model parameters and generation of the data}

The complete definition of the generative model for eye-tracking measures described above requires that we define a number of parameters: the number of subjects (\texttt{n.subjects}), the number of items (\texttt{n.items}), the standard deviations of the random effects for subjects and items (\texttt{sd.subjects}, \texttt{sd.items}), the probability of a refixation (\texttt{p.refix}), the probability of a regression (\texttt{p.regr}), the probability of rereading after the first pass (\texttt{p.reread}), the average duration of the first fixation (\texttt{mean.ffd}), the average summed duration of all first-pass refixations (\texttt{mean.gazediff})\footnote{The parameter is named \texttt{mean.gazediff} because it quantifies the mean \textit{diff}erence between first fixation duration and gaze duration.  Thus the mean gaze duration can be obtained by summing \texttt{mean.ffd} and \texttt{mean.gazediff}.  The names of the following parameters are motivated in the same way.}, the average summed duration of all fixation occurring after a regressive saccade and before anything to the right of the word is fixated (duration of the regression path) (\texttt{mean.gopastdiff}), the average summed duration of all fixations on the word occurring after the first pass (\texttt{mean.tvtdiff}), and the standard deviations of all these durations (\texttt{sd.ffd}, \texttt{sd.gazediff}, \texttt{sd.gopastdiff}, \texttt{sd.tvtdiff}).

The model samples durations from log-normal distributions which resemble the distributions found in true eye-tracking measures more closely than normal distributions.  All means and their standard deviations are therefore given as geometric means and geometric standard deviations.  Note that the geometric mean is, like the arithmetic mean, on the milliseconds scale.  However, the geometric standard deviation is dimensionless and therefore does not have a unit (unlike the ``arithmetic'' standard deviation which is measured in milliseconds).

% TODO Use past tense throughout in this section.

Once the parameters listed above are defined, the model generates eye-tracking measures in the following way: By-subject random intercepts ($u_j$, with $j$ ranging from 1 to \texttt{n.subjects}) and by-item random intercepts ($w_k$, with $k$ from 1 to \texttt{n.items}) are sampled from normal distributions with $\mu = 0$ and $\sigma = log(\texttt{sd.subjects})$ and $\sigma = log(\texttt{sd.items})$.  First fixation durations $ffd_{jk}$ are then sampled from a log-normal distribution with $\mu = e^{ln(\texttt{mean.ffd}) + u_i + w_k}$ and $\sigma = \texttt{sd.ffd}$.  Based on the parameter \texttt{p.refix}, the model determines whether or not there was a refixation in a trial.  If yes, gaze duration was generated by sampling from the log-normal distribution defined by $\mu = \texttt{mean.gazediff}$ and $\sigma = \texttt{sd.gazediff}$ and by adding the result to the corresponding first fixation duration.  If there was no refixation, the gaze duration was the same as the first fixation duration.  Go-past times and total viewing times were calculated in the same fashion.

Each virtual subject was assigned to one list of stimuli following a Latin square design as it would be done in a real experiment.  Each iteration of the simulation tested seven different versions of the data set which differed in the size of the true effect that was added to all four eye-tracking measures.  Effectively, we introduced the effect in first fixation duration and the later measures inherited it from that.  Effects were added on the log-scale such that they had the desired size on the milliseconds scale.  The effect sizes were: 0, 2.5, 5, 10, 20, 40, and 80\,ms.  The first version (without true effect) was used to determine the rate of false positives.  The other versions (with true effect) were used to investigate statistical power.

\subsection{Calculation of model parameters}
\label{calculation-of-model-parameters}

There is of course not a single set of parameter values that describes eye-tracking measures collected across different types of reading experiments.  Depending on the stimuli, the instructions, and the subject population, the parameters can rather take a range of values.  In order to increase the generalizability of our results, we therefore sampled model parameters from a range of plausible values that cover a wide range hypothetical reading experiments.  Specifically, each iteration of the Monte Carlo simulation used a unique parameter set that was sampled from a multidimensional uniform distribution whose dimensions corresponded to the parameters of the model.  The end points of this distribution were marked by the parameters found in two published reading experiments: \citeNP{AngeleEtAl2013} and \citeNP{MetznerEtAl2016}.

These experiments were chosen because, first, they represent the class of experiments that we are interested in (standard reading experiments) and, second, because their data differed significantly in relevant ways: The experiment by Angele and colleagues investigated parafoveal preview.  It had low rates of refixations and regressions because the sentences were easy to read and the manipulation (type of preview shown for the target word) was relative subtle.  The eye-tracking measures therefore showed relatively high correlations.  In contrast to that, the experiment by Metzner and colleagues used difficult sentences that included syntactic or semantic violations in four out of overall six conditions.  As a result, there were high rates of refixations and regressions in the target region, which led to relatively low correlations of eye-tracking measures.  Thus, we expected a smaller increase in false positives due to multiple testing for data sets resembling the study by Angele et al. than for data sets resembling the study by Metzner et al.

The numbers of subjects and items used in the studies by Angele et al. and Metzner et al. were relatively high because both studies had more complex designs than is assumed in the present simulations.  We therefore used numbers of subjects and items that are more representative for experiments with simple designs, ranging from 20 to 50 subjects and items.

Table~\ref{parameter-ranges} shows the ranges of parameters that were used in the simulation.

\begin{table}[h]
\caption{The upper and lower bounds of the parameter ranges used to generate the data.  The lower bound values are based on the data reported in Angele et al. (2013) and the upper bound values are based on Metzner et al. (2016).  The exception are \texttt{n.subjects} and \texttt{n.items} which were chosen to reflect the range typically found for experiments with simple designs.}
\centering

\begin{tabular}{lrr}
\toprule
Parameter & Lower bound & Upper bound\\
\midrule
n.subjects & 20.00 & 50.00\\
n.items & 20.00 & 50.00\\
sd.subjects & 1.10 & 1.16\\
sd.items & 1.07 & 1.08\\
\addlinespace
p.refix & 0.14 & 0.32\\
p.regr & 0.07 & 0.43\\
p.reread & 0.19 & 0.41\\
\addlinespace
mean.ffd & 219.72 & 232.35\\
mean.gazediff & 197.09 & 204.27\\
mean.gopastdiff & 312.02 & 558.27\\
mean.tvtdiff & 242.26 & 291.40\\
\addlinespace
sd.ffd & 1.31 & 1.43\\
sd.gazediff & 1.40 & 1.69\\
sd.gopastdiff & 1.69 & 1.84\\
sd.tvtdiff & 1.55 & 1.87\\
\bottomrule
\end{tabular}

\label{parameter-ranges}
\end{table}

\subsection{Evaluation of artificial data}
\label{evaluation-of-artificial-data}

For the purpose of this study, it is important that the artificial data exhibit similar variance components and correlations among eye-tracking measures as found in real data.  Below, we compare the data reported in Angele et al. and the Metzner et al. study with artificial data generated to resemble these data sets.

We first calculated the pair-wise correlations of the four eye-tracking measures in the real data and then in the artificial data. Table~\ref{table:correlations} shows that the overall pattern of correlations was largely preserved in the artificial data.  The exception are the correlations between go-past duration (GPD) and total viewing time (TVT) which were too low in the artificial data.  Apparently, there is a connection between these measures that our model does not capture.

The correlations were generally higher in the Angele et al. data than in the Metzner et al. data and this is also reflected in the artificial data sets.  We also see that correlations tend to be somewhat higher in the artificial data than in the real data.  This is a consequence of the simplifying assumptions made in our model which has fewer sources of variance than we can realistically assume to be the present in the process generating the real data.  A consequence of this is that our simulations may slightly underestimate the rate of false positives.

\begin{table}
\caption{Correlations of eye-tracking measures in real data (left) and in artificial data modeled after the real data (right)}
\centering
\begin{tabular}{lrrrrcrrrr}
\toprule
 & \multicolumn{4}{c}{Angele et al. data} & & \multicolumn{4}{c}{Artificial data}\\
\midrule
 & FFD & GZD & GPD & TVT & & FFD & GZD & GPD & TVT\\
\addlinespace
FFD & 1.00 & 0.64 & 0.40 & 0.40 & & 1.00 & 0.65 & 0.46 & 0.42\\\addlinespace
GZD & 0.64 & 1.00 & 0.63 & 0.62 & & 0.65 & 1.00 & 0.71 & 0.64\\\addlinespace
GPD & 0.40 & 0.63 & 1.00 & 0.61 & & 0.46 & 0.71 & 1.00 & 0.45\\\addlinespace
TVT & 0.40 & 0.62 & 0.61 & 1.00 & & 0.42 & 0.64 & 0.45 & 1.00\\\addlinespace
\bottomrule
\addlinespace
 & \multicolumn{4}{c}{Metzner et al. data} & & \multicolumn{4}{c}{Artificial data}\\
\midrule
 & FFD & GZD & GPD & TVT & & FFD & GZD & GPD & TVT\\
\addlinespace
FFD & 1.00 & 0.50 & 0.11 & 0.24 & & 1.00 & 0.61 & 0.20 & 0.36\\\addlinespace
GZD & 0.50 & 1.00 & 0.24 & 0.52 & & 0.61 & 1.00 & 0.35 & 0.57\\\addlinespace
GPD & 0.11 & 0.24 & 1.00 & 0.49 & & 0.20 & 0.35 & 1.00 & 0.19\\\addlinespace
TVT & 0.24 & 0.52 & 0.49 & 1.00 & & 0.36 & 0.57 & 0.19 & 1.00\\\addlinespace
\bottomrule
\end{tabular}
\label{table:correlations}
\end{table}

\begin{figure}
\centering
\begin{knitrout}
\definecolor{shadecolor}{rgb}{0.969, 0.969, 0.969}\color{fgcolor}
\includegraphics[width=\maxwidth]{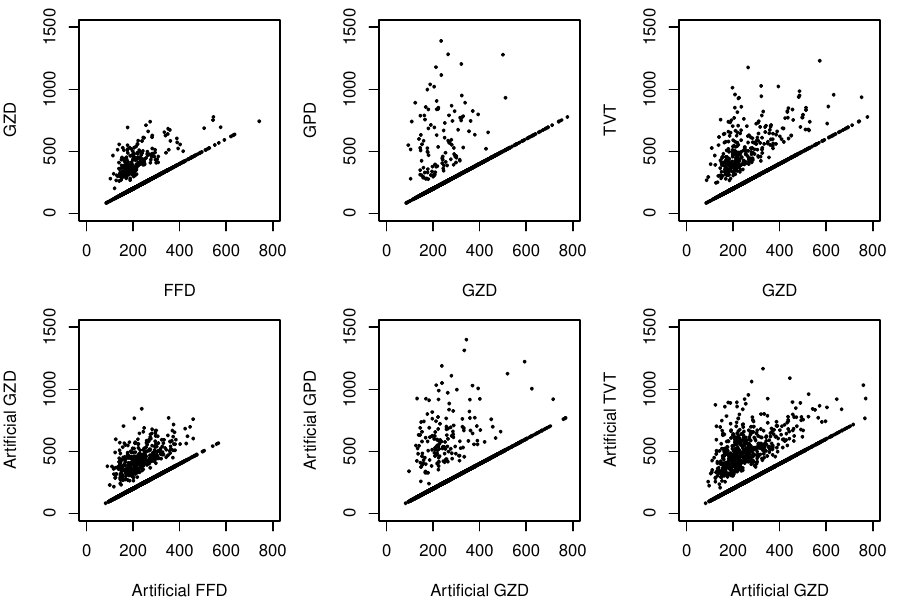}

\end{knitrout}
\caption{Scatter plots showing the relationships among measures for the Angele et al. data: GZD as a function of FFD (left), GPD as a function of GZD (middle), TVT as a function of GZD (right).  The first row shows the real measures, the second the artificial measures.  A subset of 1600 randomly selected data points (\textasciitilde 50\%) were used in this plot to improve readability.}
\label{figure_scatter_plots_angele}
\end{figure}

\begin{figure}
\centering
\begin{knitrout}
\definecolor{shadecolor}{rgb}{0.969, 0.969, 0.969}\color{fgcolor}
\includegraphics[width=\maxwidth]{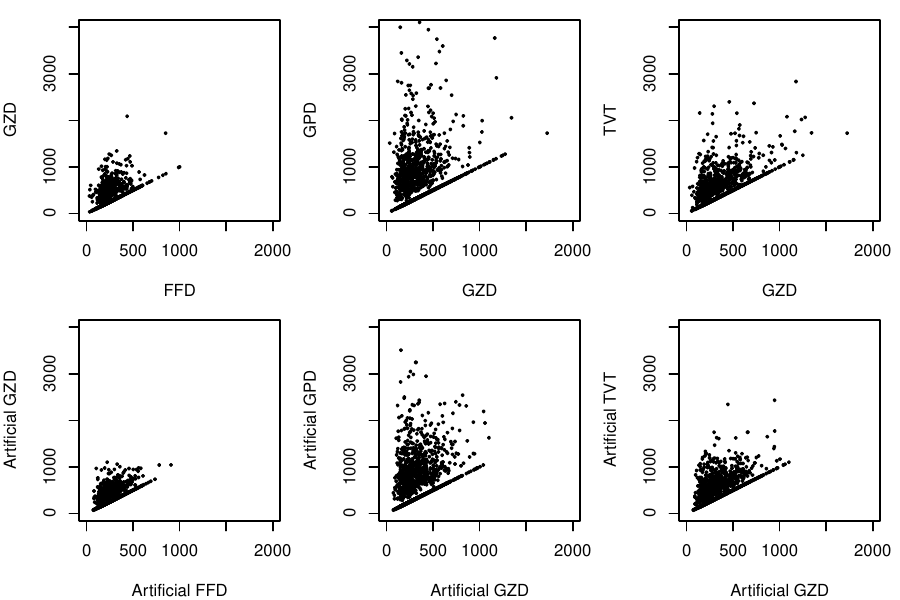}

\end{knitrout}
\caption{Scatter plots showing the relationships among measures for the Metzner et al. data: GZD as a function of FFD (left), GPD as a function of GZD (middle), TVT as a function of GZD (right).  The first row shows the real measures, the second the artificial measures.  A subset of 1600 randomly selected data points (\textasciitilde 10\%) were used in this plot to improve readability.}
\label{figure_scatter_plots_metzner}
\end{figure}

A more detailed picture of the dependencies between eye-tracking measures can be obtained from scatter plots that show each measure as a function of another measure.  Fig.~\ref{figure_scatter_plots_angele} and Fig.~\ref{figure_scatter_plots_metzner} show that these dependencies are faithfully reproduced by our model.

\subsection{Monte Carlo simulation}
\label{monte-carlo-simulation}

Having established that our artificial data is a sufficiently good approximation of real eye-tracking data, we can proceed to the discussion of the Monte Carlo simulation.  This simulation consisted of 100,000 iterations.  In each iteration we generated seven data sets, one for each tested effect size (0--80\,ms), and tested each of these data sets with the three decision criteria.

We fit one linear mixed model for each dependent measure \cite<FFD, GZD, GPD, TVT;>[]{BatesEtAl2015}.  All dependent measures were log-transformed to obtain normally distributed residuals.  The models had random intercepts for subjects and items.  Random slopes were not needed because we generated the data such that random slopes would not systematically vary across subjects and items.  The \emph{p}-values for the effect of the manipulation were determined by fitting models with and without the factor for this manipulation and by comparing these models using likelihood-ratio tests \cite{BatesEtAl2015}.

For each linear mixed model, we also recorded whether or not there was a convergence warning.  No convergence warnings occurred in the overall 1,400,000 linear mixed-models.\footnote{Due to the large number of statistical tests, we ran the simulation on a Slurm cluster \cite{YooEtAl2003} with 110 cores (including hyperthreading cores) made available by four compute servers running Ubuntu Linux.}

The purpose of the simulation was to empirically investigate false positive rates and statistical power under the three decision criteria discussed in the introduction. Any significant effect of the hypothetical manipulation found in a data set that did not have a true effect was a false positive by definition.  If a significant effect was found in such a data set, that result is a Type I error caused by random variation in the data.  Thus, we simply had to count the number of significant effects found under each decision criterion and assess whether the rate of false positives was higher or lower than the conventionally accepted 5\%.

Similarly, we assessed statistical power using the data sets that did contain a true effect.  Using each of the three decision criteria we tested whether an effect was present or not in each data set and the rate of successfully detected effects then constituted our measure of power.

\section{Results}
\label{results}

\begin{table}
\caption{Rates of false positives under the three decision criteria and binomial 95\% confidence intervals}
\centering

\begin{tabular}{lrr}
\toprule
Criterion & False positive rate & 95\% Confidence interval\\
\midrule
One effect & 14.8\% & 14.6--15.1\%\\
Two effects & 3.9\% & 3.8--4.1\%\\
Bonferroni & 4.2\% & 4.1--4.3\%\\
\bottomrule
\end{tabular}
\label{table:false-positives}
\end{table}

\subsection{False positives under the three decision criteria}

Table~\ref{table:false-positives} shows the false positive rates obtained under each of the three decision criteria.  Applying the baseline criterion, which demands only one effect with $p\leq0.05$ (no correction for multiple comparisons), led to a false positive rate of 15\%. This rate is lower than the 18.5\% we would expect if the tests were statistically independent ($1-0.95^4=0.185$) but not by much.  Clearly, this rate is much higher than the conventionally accepted 5\%.  A failure to correct for multiple comparisons will therefore result in an unacceptably high risk that spurious effects will be found even if there is no real effect.

As expected, the Bonferroni correction is an effective remedy against inflated false positives.  Under the Bonferroni correction, we obtained false positives in 4.2\% of the cases.  Although this is slightly more conservative than necessary, it is fairly close to the desired rate of 5\%.

Requiring effects with $p\leq0.05$ in at least two eye-tracking measures (rule-of-thumb criterion), produced a false positive rate of 3.9\% which is also close to the desired 5\% but even more conservative than the Bonferroni correction.

A preliminary conclusion from these results is that the multiple comparisons problem cannot be ignored and that both alternative decision criteria help to keep false positives under control.

\subsection{Factors influencing the rate of false positives}

The results reported above were obtained by aggregating false positives from data sets with a wide range of different parameters.  Crucially these data sets differed with respect to parameters that influence the correlations among eye-tracking measures which in turn influence false positive rates.  Thus, a more detailed picture can be obtained, by investigating how these parameters influence false positives rates.  This exercise also sheds light on the extent to which the results reported in Table~\ref{table:false-positives} generalize to a wide range of experiments.

Specifically, we tested the effects of the parameters for the probability of refixations (\texttt{p.refix}), regressions (\texttt{p.regr}), and rereadings (\texttt{p.reread}), and the average durations of initial fixations (\texttt{mean.ffd}), refixations (\texttt{mean.gazediff}), regression paths (\texttt{mean.gopastdiff}), and second-pass refixations (\texttt{mean.tvtdiff}).  If these parameters have high values, the correlations among measures are reduced and thus the chance of finding false positive effects is increased.  The question is by how much.

\begin{figure}
\centering
\begin{knitrout}
\definecolor{shadecolor}{rgb}{0.969, 0.969, 0.969}\color{fgcolor}
\includegraphics[width=\maxwidth]{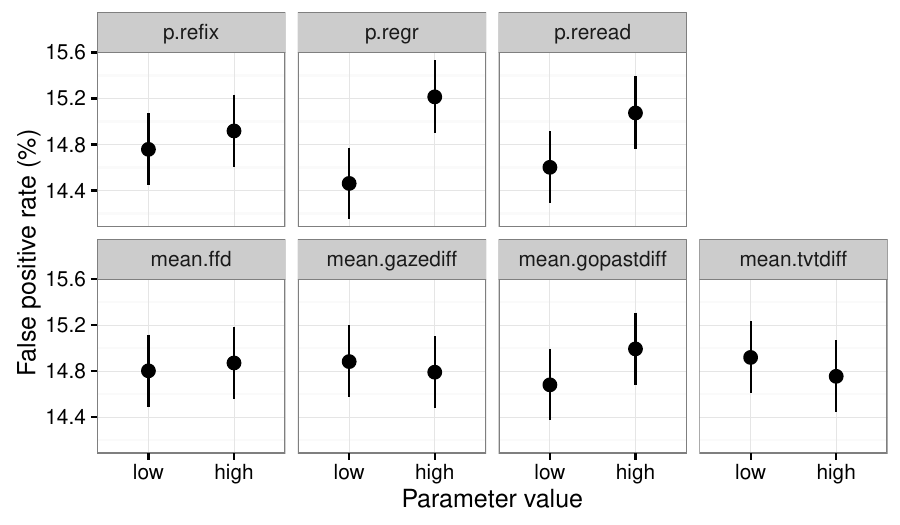}

\end{knitrout}
\caption{Influence of parameter values on the rate of false positives when no correction is applied.  ``Low'' means that the parameter had values closer to those found in the experiment by Angele et al., ``high'' means the parameter values were closer to those found in Metzner et al.} 
\label{figure_false-positives-by-parameters}
\end{figure}

Fig.~\ref{figure_false-positives-by-parameters} shows for each parameter the difference in false positive rates between data sets with a high and a low parameter value (median split).  The most important thing to be gleaned from this graph is that each parameter on its own had a relatively minor influence on the rate of false positives ($\leq0.8\%$).  The parameters \texttt{p.refix}, \texttt{mean.ffd}, \texttt{mean.gazediff}, and \texttt{mean.tvtdiff} had no reliable effect on false positives, and the parameters \texttt{p.regr}, \texttt{p.reread}, and \texttt{p.gopastdiff} had robust but small effects in the expected direction.

When these parameters conspire, these smaller effects can add up to produce a slightly larger difference.  Table~\ref{table:false-positives-max-effect} shows the rates of false positives for data sets with high values for parameters that showed a significant effect compared to the rates found in data sets with low parameter values.  In other words, this graph shows false positive rates for data sets resembling the experiment by Metzner et al. (15.7\%) to those resembling the experiment by Angele et al. (14.0\%).  The difference between the rates of false positives was only 1.7\%, which suggests that the main result -- unacceptably high rates of false positives when no correction is applied -- holds for a wide range of different data sets. 

\begin{table}
\caption{Rates of false positives when applying no correction for data sets resembling the Metzner et al. experiment (high parameter values, low correlations among eye-tracking measures) and data sets resembling Angele et al.'s study (low parameter values, high correlations among eye-tracking measures)}
\centering
\begin{tabular}{lrrr}
\toprule
Type of data set & False positive rate & 95\% Confidence interval\\
\midrule
Resembling Metzner et al. & 15.7\% & 15.1--16.4\%\\
Resembling Angele et al. & 14.0\% & 13.4--14.6\%\\
\bottomrule
\end{tabular}
\label{table:false-positives-max-effect}
\end{table}

\subsection{Statistical power}

The results above show that corrections for multiple comparisons are necessary even if only four eye-tracking measures are tested for effects of a manipulation.  Both corrections, the Bonferroni correction and the rule-of-thumb criterion, help to keep false positives down to conventionally acceptable levels.  However, this inevitably comes at the cost of reduced statistical power.  Since a researcher planning a study needs to take statistical power into account, for example, when deciding about the number of participants and experimental items, it is important to understand how much power is sacrificed by applying corrections for multiple testing.

As described above, we generated data sets with true effect sizes of 0, 2.5, 5, 10, 20, 40, and 80\,ms.  Detecting effects of 2.5\,ms was expected to be difficult with all three decision criteria, whereas effects of 80\,ms were expected to be easy to detect.  Beyond that, we expected the three decision strategies to differ in how reliably they detect effects between these extremes.

\begin{figure}
\centering
\begin{knitrout}
\definecolor{shadecolor}{rgb}{0.969, 0.969, 0.969}\color{fgcolor}
\includegraphics[width=\maxwidth]{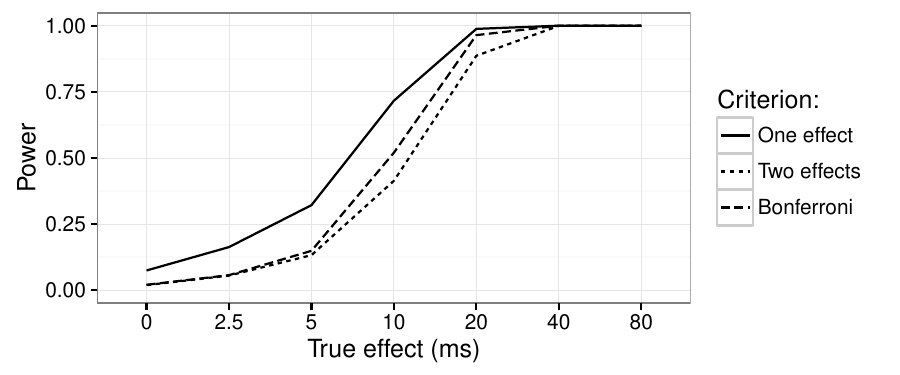}

\end{knitrout}

\caption{Statistical power as a function of effect size for the three tested decision criteria.  At 0\,ms the graph shows the power that we obtain in the limit as the effect size approaches zero.} 
\label{figure_power_by_criterion}
\end{figure}

Fig.~\ref{figure_power_by_criterion} shows the detection rates for all three decision criteria as a function of effect size. For effect sizes greater than zero, the detection rate indicates the proportion of correctly detected true effects of the hypothetical manipulation.  By ``correctly detected'' we mean that it was not enough that there was a significant effect, the effect also had to be in the correct direction.  For effect size zero, the graph shows the detection rate that we obtain in the limit as the effect size approaches zero.\footnote{Note that this detection rate is not the false positive rate.  False positives are significant effects in the correct or in the incorrect direction but the detection rate only counts effects in the correct direction.  This means that the detection rate at 0\,ms is half the false positive rate because significant effects are equally likely to go in either direction at effect sizes approaching zero.} 

Under all three decision criteria (one significant effect, two significant effects, one significant effect after Bonferroni correction) we had difficulties detecting effects of 5\,ms and below.  All criteria also detected effects of 20\,ms and larger with near perfect accuracy.  For effects between 5 and 20\,ms, we see that the baseline criterion (one effect, no correction) has the highest power, but, as we now know, this comes at the price of a greatly elevated rate of false positives.

Under the Bonferroni correction, power is reduced.  However, the reduction in power is relatively small; less than \textasciitilde 20\%.  This is surprising because the conventional wisdom and main argument against using the Bonferroni correction is that the Bonferroni correction massively reduces power, such that experiments have no reasonable chance to detect true effects.  This is clearly not true and our results suggest that only a moderate amount of statistical power needs to be sacrificed to control false positives.

The rule-of-thumb criterion (two significant effects) led to a loss in power of up to \textasciitilde 30\% compared to the baseline criterion.  Compared to the Bonferroni correction, the rule-of-thumb criterion performed particularly poorly at effect sizes of 10 and 20\,ms and was otherwise similar to the Bonferroni correction.

\subsection{Power when only a subset of measures is affected by the manipulation}

\begin{figure}
\centering
\begin{knitrout}
\definecolor{shadecolor}{rgb}{0.969, 0.969, 0.969}\color{fgcolor}
\includegraphics[width=\maxwidth]{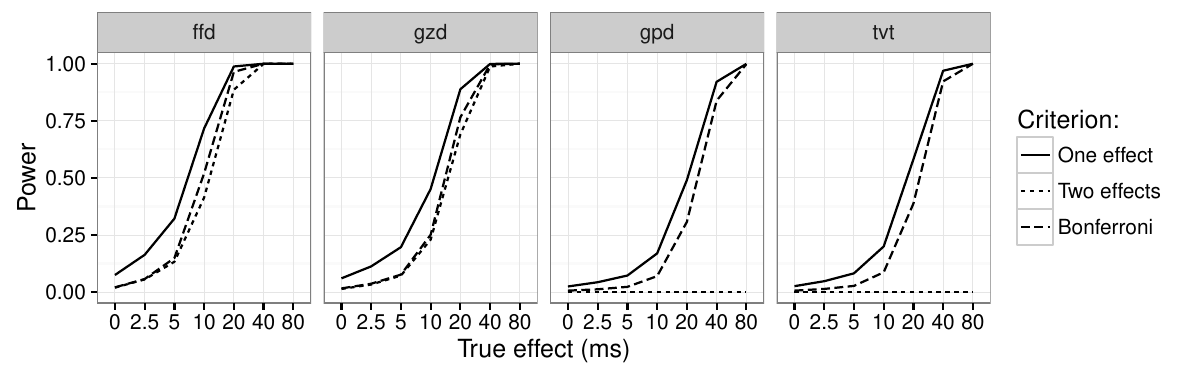}

\end{knitrout}

\caption{This figure shows how power changes when the true effect is introduced at later stages, i.e., when only a subset of measures are affected by the manipulation.  Statistical power is shown as a function of effect size for the three tested decision criteria.  The first panel shows the results for when the effect is introduced in first fixation duration (same as Fig.~\ref{figure_power_by_criterion}, repeated here for convenience).  The second panel shows results for when the effect is introduced in refixations.  The third panel is for the case where only go-past time has a true effect, i.e. the effect occurred during regressions.  The fourth panel is for total viewing time, i.e. when the effect is introduced in rereading times.  At 0\,ms the graph shows the power that we obtain in the limit as the effect size approaches zero.} 
\label{figure_power_by_stage}
\end{figure}

In the simulations above, the true effect was introduced during the first fixation.  This means that the effect is present in all four measures because they all contain the duration of the first fixation as a component.  While this is a common scenario, it is also possible that a true effect occurs at a later stage of processing, for example during refixations.  In this case, only a subset of the measures can possibly show the effect, which makes successful detection less likely.  Is the Bonferroni correction still appropriate in these cases?  To answer this question we conducted three additional simulations, one in which the effect was introduced in refixations (true effect in GZD, GPG, and TVT), one in which it was introduced during regressions (true effect only in GPG), and one in which the effect was introduced in rereading times (true effect present only in TVT).  The results are shown in Fig.~\ref{figure_power_by_stage}.

In all three scenarios, there is only a moderate reduction in power due to the Bonferroni correction and this reduction is similar to that observed when the effect is present in all four measures.  Power is reduced considerably when only a subset of measures have a true effect but that lies in the nature of things and is true independently of whether or not the Bonferroni correction is used.  Further, we see that the rule-of-thumb criterion fails catastrophically when the true effect is present in only one measure.  This is of course unsurprising because it directly follows from the definition of that criterion.

\subsection{Power needed to detect effects in individual measures}

In many cases, the detection of an effect in any of the measures is enough to support a tested hypothesis.  However, it is also quite common to draw conclusions based on the identity of the measures in which an effect was detected.  For example if an effect appears in a so-called early measure (e.g., first fixation duration), researchers may conclude that the manipulation affected rapid cognitive processes whereas an effect in a so-called late measure may be attributed to temporally delayed processes \cite<see>[for a discussion of early and late effects]{CliftonEtAl2007}.  \citeNP{VasishthEtAl2013} argue that this reasoning is not perfectly sound due to the way how these measures are defined: most so-called late measures (e.g, total viewing time) are composed of early fixations and late fixations and may therefore reflect early and late processes.  However, there is another issue that complicates inferences about the time-course of cognitive processes and this issue has to do with statistical power.

Early and late measures have different variances and this means that effects are harder to detect in some measures than in others.  Specifically, early measures like first fixation duration tend to have lower variance than later measures like rereading time (summed duration of fixations occurring after the first pass).  So if an effect appears in first fixation duration but not in rereading time that could either mean that the effect was absent in rereading time but it could also mean that it was harder to detect in rereading time due to lower statistical power for that measure.

\begin{figure}
\centering
\begin{knitrout}
\definecolor{shadecolor}{rgb}{0.969, 0.969, 0.969}\color{fgcolor}
\includegraphics[width=\maxwidth]{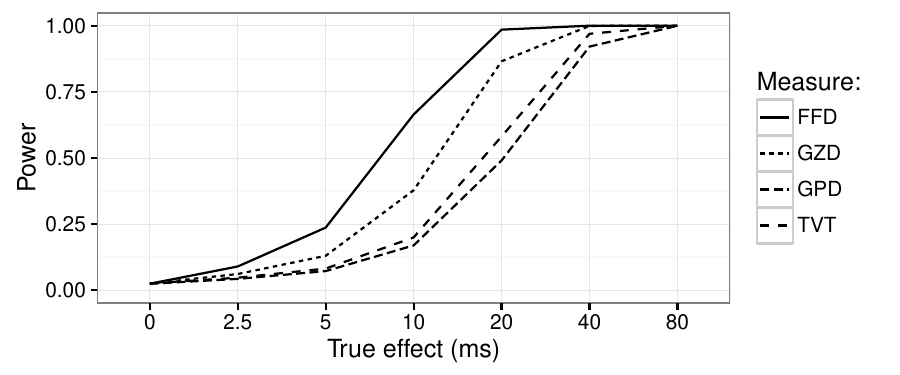}

\end{knitrout}
\caption{Power as a function of true effect size for all four eye-tracking measures separately.  The measures were first fixation duration (FFD), gaze duration (GZD), go-past time (GPD), and total viewing time (TVT).  No correction for multiple testing was applied.  At 0\,ms the graphs show the power that we obtain in the limit as the effect size approaches zero.} 
\label{figure_power_by_measure}
\end{figure}

Our simulations allowed us to quantify these differences in power between eye-tracking measures and thus helps us to get a sense of how much of a problem differences in power might be.  Fig.~\ref{figure_power_by_measure} shows the rate of correctly detected effects for each measure as a function of the size of the true effect.  We can see that the differences in power were quite dramatic.  While an effect of 20\,ms was reliably detected in first fixation duration, go-past time showed the effect only in about 50\% of the cases.  We also see that reliable detection requires effects four times bigger in go-past time and total viewing time than in first fixation duration.  These results suggest that the pattern of effects across measures is strongly influenced by differences in power.  Hence, special care is needed when interpreting the identity of the measures in which an effect occurs.

\subsection{Boosting statistical power}

\begin{figure}
\centering

\includegraphics[width=\maxwidth]{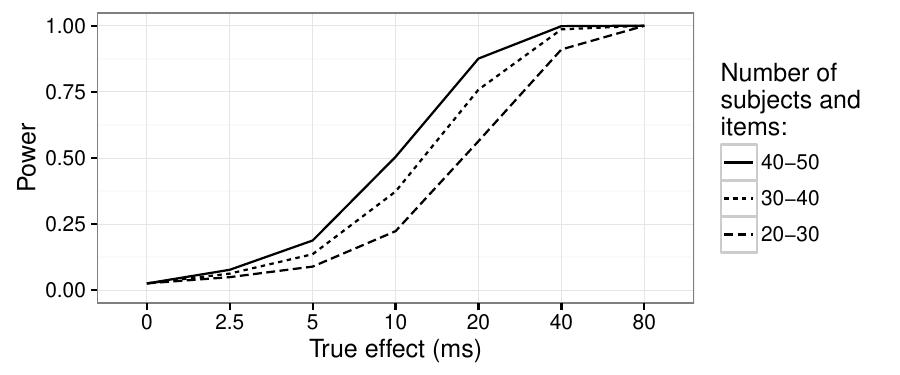}

\caption{Power as a function of true effect size and the number of subjects and items.  At 0\,ms the graphs show the power that we obtain in the limit as the effect size approaches zero.} 
\label{figure_power_by_subjects_and_items}
\end{figure}

Most factors influencing statistical power are not under the control of the researcher and thus it is not entirely trivial to offset the loss in power due to corrections for multiple testing.  Figures~\ref{figure_power_by_criterion} and \ref{figure_power_by_measure} show that increasing the effect size has a strong impact on power but the means for doing so are usually very limited.  The only relevant parameters that a researcher can fully control are the numbers of subjects (\texttt{n.subjects}) and experimental items (\texttt{n.items}).  Fig.~\ref{figure_power_by_subjects_and_items} shows how statistical power depends on these numbers.

Increasing the number of subjects and items from \textasciitilde 25 to \textasciitilde 45 resulted a gain in power of up to \textasciitilde 31\%.  Subjects and items contributed equally to this effect (based on plots not shown in this manuscript).  This gain is more than enough to compensate for the power lost when applying the Bonferroni correction (less than \textasciitilde 20\%).

\subsection{Consequences of low power: Type S and Type M errors}

Although we can in principle test arbitrarily many subjects and items, there are usually resource limitations that prevent us from doing so.  Hence, a researcher has to make a careful decision about the number of subjects and items that results in high-enough power but does not waste resources.  This decision must be made in advance, i.e., before data collection begins \cite{Kruschke2010, Wagenmakers2007}, and ideally with the aid of a power analysis \cite{VasishthNicenboim2016}.\footnote{Ben Bolker published a tutorial in which he shows how to run power analyses for designs with repeated measurements for subjects and items: \url{https://rpubs.com/bbolker/11703}}

When making this decision, it is of course important to fully understand the statistical consequences of running studies with too little power.  One obvious consequence of low power is that we may not detect a true effect (Type II error).  However, even if we do detect an effect, power is still an important factor that needs to be considered when interpreting that effect.  An extreme case arises when power is close to zero.  Any effect found in this situation is more likely caused by random variation than by the investigated cognitive processes.  A researcher should therefore not base any conclusions on this effect even if it is highly significant.\footnote{P-values are uniformly distributed between 0 and 1 when the null-hypothesis is true.  Hence, a very small p-value is just as likely as any other p-value when there is no real effect.}  However, even when power was considerably higher than zero, we may still fall victim to two types of statistical errors that are not widely know in the field but which can have catastrophic consequences for theory development.

\begin{figure}
\centering

\includegraphics[width=\maxwidth]{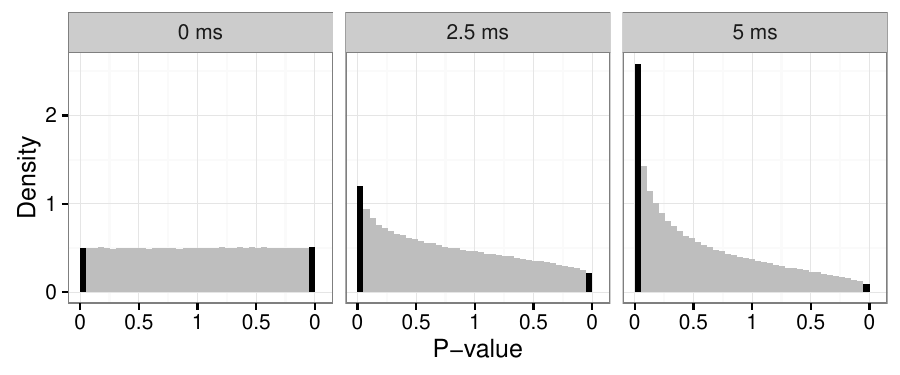}

\caption{Histograms of p-values for effect sizes of 0, 2.5, and 5\,ms.  The left half of each panel show the p-values for observed differences in the correct direction, the right half, on a mirrored scale, the p-values for observed differences in the wrong direction.  Black bars denote the bins with p-values $\leq0.05$, i.e. significant effects.  The panel titled ``0\,ms'' shows the histogram that we get in the limit as the true effect size approaches zero.}
\label{figure_type-s-errors}
\end{figure}

\citeNP{GelmanCarlin2014} discuss these two errors types which they call Type S and Type M errors.  A Type S (\textit{sign}) error is an error where a significant effect is detected but the direction of the detected effect is the opposite of the true effect.  If we have high power, for example because the effect is large, this is unlikely to happen.  However, if power is low, random variation may outweight the true effect and create an observed effect in the opposite direction.

Fig.~\ref{figure_type-s-errors} shows histograms of p-values for differences between conditions in the correct direction and in the incorrect direction.  Bins with p-values $\leq0.05$ (i.e., significant effects) are black, bins for p-values $>0.05$ (non-significant differences) are grey.

For true effect sizes approaching zero (left panel), we see that significant effects in the correct direction were as likely as significant effects in the incorrect direction.  This is expected because there is really no influence of the vanishingly small true effect, so any significant effect must be due to random variation which can go in either direction.

For a true effect size of 2.5\,ms (middle panel), we see that the true effect exerted some force and increased the rate of correctly detected effects.  However, random variation can still outweight the true effect and create significant observed effects in the wrong direction (Type S error).  Specifically, there was a 15\% chance that a significant effect was in the wrong direction.  Given that this kind of error has catastrophic consequences for theory development, this is a fairly unsettling number.  One might say that effects this small are not relevant, but the literature on parafoveal preview has several reports of effects that are nearly as small as 2.5\,ms (e.g., \cite{RisseKliegl2011}; see \cite{VasilevAngele2016Psychonomics} for a review).

For true effects of 5\,ms (right panel), the risk of Type S errors diminished further due to improved power.  However, there was still a 3.4\% chance that a significant effect was in the wrong direction.

\begin{figure}
\centering

\includegraphics[width=\maxwidth]{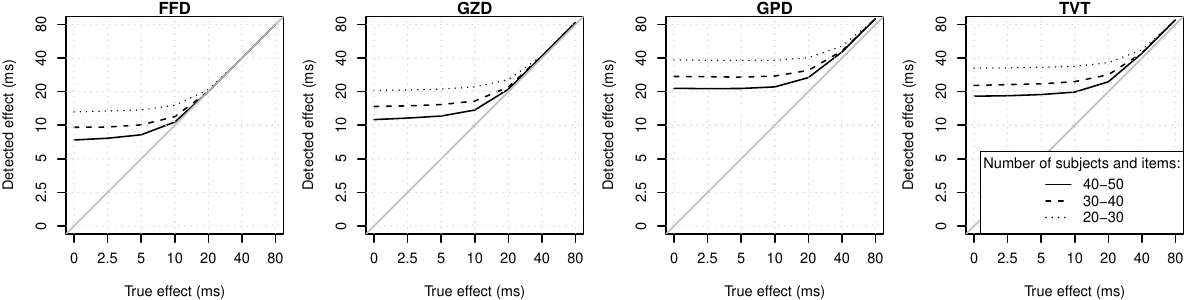}

\caption{Size of detected effects as a function of true effect size for each eye-tracking measure and for different numbers of subjects and items.  The grey line shows the result that we would expect from a (hypothetical) statistical test that is immune to Type M errors.}
\label{figure_type-m-errors}
\end{figure}

Type M (\textit{magnitude}) errors arise in situations where an effect in the correct direction is detected but the size of the true effect is overestimated.  Again, this is more likely to happen -- more precisely: the error tends to be larger -- when power is low.  If the power is not sufficient to reliably detect the true effect, it may emerge as significant only when random variation amplifies the true effect such that it can pass the threshold for significance.  Hence, the detected effect size will be larger than the true effect size in these cases.  In situations where effect size matters, an overestimated effect can of course lead to false conclusions.

Fig.~\ref{figure_type-m-errors} shows the size of detected effects as a function of the size of the true effect, for different numbers of subjects and items, and for each eye-tracking measure separately.  One observation is that the difference between conditions needed to have a certain size (\textasciitilde 7--40\,ms) to pass the threshold for significance and that this difference was often larger than the true effect size.  The result is that the effect size was substantially overestimated for all true effect sizes below that threshold.  Even above that threshold, the effect size was overestimated because an additional boost from random variation made it more likely that a true difference turned out significant.

The size of Type M errors was different depending on the tested eye-tracking measure.  First fixation duration, which, as we now know, provides the highest power among the measures investigated here (see Table~\ref{figure_power_by_measure}), had relatively small Type M errors, while go-past duration showed relatively large Type M errors as a consequence of low power.  Finally, we see that the size of the Type M errors was approximately twice as big in data sets with low numbers of subjects and items than in data sets with high numbers of subject and items.

\section{Discussion}
\label{discussion}

The primary purpose of the simulations reported above was to answer two questions.  First, we wanted to determine the rate of false positive results obtained when four dependent eye-tracking measures are tested for the effect of an experimental manipulation and when no correction for multiple comparisons is applied.  Second, we wanted to determine to what extent two alternative decision criteria are effective at keeping false positives under control while preserving as much statistical power as possible.

Our results are quite clear: When four dependent eye-tracking measures are tested without correction for multiple comparisons, the rate of false positive results is unacceptably high (14.8\%).  A failure to address multiple comparisons may therefore considerably compromise the replicability of a result, and researchers and reviewers should be aware of this problem.

Both alternative decision criteria were slightly too conservative but kept the rate of false positives close to the conventionally accepted 5\%.  However, the Bonferroni correction preserved more statistical power and is therefore preferred.  In the following, we will discuss the alternative decision criteria in more detail.

\subsection{Bonferroni correction}
\label{bonferroni-correction}

The Bonferroni correction was slightly more conservative than required but statistical power was only moderately reduced.  Contrary to conventional wisdom, the Bonferroni correction is therefore quite appropriate when multiple dependent eye-tracking measures are tested.  Since the Bonferroni correction was slightly too conservative, a researcher applying this correction may in some cases not find evidence for a true effect when a more precise correction may have allowed the detection of the effect.  However, this type of error (Type II, false negative) has a much smaller potential to do harm than the greatly inflated false positive rate suffered without any form of correction.  After all, researchers will typically draw strong conclusions when they believe to have found statistical evidence for an effect whereas a null-result (no statistical evidence for an effect) is inconclusive.

A researcher reporting an effect in just one dependent variable may argue that a Bonferroni correction is not needed because the effect was expected to appear in that dependent variable and not in the others.  If that were the case, the reader of such research may rightly ask, why did the researchers test the other dependent variables in the first place?  Even if the tested hypothesis predicts an effect in just one particular measure, it is often possible to produce a perfectly plausible explanation for why the effect could also appear in another dependent measure.  Clearly, our general tendency to view an effect as entirely plausible after seeing the evidence for it does not help here.  Thus, a researcher either has to demonstrate the a-priori nature of the prediction beyond reasonable doubt (e.g., through preregistration of the study design or through a very strong, independently motivated hypothesis), or correct for multiple testing, or conduct a replication of the experiment \cite<see>[for further discussion]{GelmanLokenMS}.

\subsection{Rule-of-thumb criterion}
\label{rule-of-thumb-criterion}

The rule-of-thumb criterion evaluated in our study required that at least two dependent variables show an effect with $p\leq0.05$.  This criterion has no rigorous mathematical foundation.  It is rather a simple and intuitive ad-hoc criterion that some researchers may use in informal settings.  The results of our simulation suggest that this criterion can perform reasonably well: its false positive rate was similar to that obtained with the Bonferroni correction but it preserved less power.  Nevertheless, there are reasons for why the more principled Bonferroni correction is clearly preferable.

One problem is that the rule-of-thumb criterion is not sensitive to the number of dependent variables that are tested.  Requiring at least two effects with $p\leq0.05$ is a strict criterion when only two dependent variables are tested, but it is not sufficient when twenty dependent variables are tested.  The latter situation is not uncommon in reading research, because researchers often evaluate a number of common eye-tracking measures calculated for several regions of interest, e.g., the pre-target region, target, and post-target region, or, in experiments using the boundary paradigm, the pre-boundary region and the post-boundary region.  Thus, it was a mere coincidence that the rule-of-thumb criterion performed so well in our simulation.  As mentioned earlier, another problem with the rule-of-thumb criterion is that it fails completely in situations where the effect can only appear in one measure.  In sum, this illustrates why informal criteria should not be used as a substitute for rigorous alternatives such as the Bonferroni correction.

\subsection{Alternative correction procedures}

Apart from the Bonferroni correction there are a number of alternative correction procedures that aim to preserve more statistical power.  Some of them do so by making additional assumptions that do not hold in the current context, but others can safely be used as drop-in replacements for the Bonferroni correction.  The Holm-Bonferroni correction, for instance, preserves uniformly more power than the Bonferroni correction \cite{Holm1979}.  However, in the simulations reported above, the gain in power due to the Holm-Bonferroni correction was never more then 1\% and we therefore decided to show results only for the simpler and more well-known Bonferroni correction.

\subsection{Rate of false positives when multiple regions are tested}

Our simulation assumes that only one region of interest is analyzed.  However, it is quite common to analyze multiple regions, e.g., the pre-target region, the target region, and one or more post-target regions.  Testing multiple regions further increases the number of statistical tests and thereby also the rate of false positives obtained when no corrections are applied.  If we make the simplifying assumption of no correlation of eye-tracking measures across regions, we can calculate by how much the false positive rate would be increased.  If we test four measures in two regions, we get a false positive rate of:  $1 - (1 - 0.148)^2 = 0.274 = 27.4\%$ ($0.148$ is the probability of a false positive if just one region is tested, see Table~\ref{table:false-positives}).  If we test three regions, we get $1 - (1 - 0.148)^3 = 0.382 = 38.2\%$.  Thus, the rate of false positives steeply increases when we test multiple regions.  The Bonferroni correction should therefore also be used to correct for tests in multiple regions of interest.

\subsection{The underappreciated role of statistical power}

Hardly any study in the field mentions statistical power although power has important implications for the inferences that we draw from the data.  For example, knowledge about the power of a statistical test allows us to tell whether a null result (failure to reject the null) is meaningfully interpretable as the absence of an effect.  If the power was high, we can plausibly argue that the failure to reject the null constitutes evidence that the null is true or at least that the effect was much smaller than hypothesized.  However, if power was low, the failure to reject the null more likely means that the data is simply inconclusive.  Yet it is almost unheard of that researchers report the power of their experiments (the present authors are no exception).

Our simulation illustrates the important role that power plays not just for Type II errors but also for Type S and Type M errors \cite{GelmanCarlin2014}.  If power is low, there is a substantial chance that a detected effect is actually in the opposite direction of the true effect, which would have catastrophic consequences for theory development.  Similarly, low power can lead to an overestimation of an effect’s size.  Our simulation also showed that power is higher for early eye-tracking measures than for late measures.  These differences in power give rise to three surprising statistical illusions:  First, higher power for early measures means that effects are more likely to emerge in those measures creating the illusion that manipulations tend to have early effects but no late effects.  Second, the lower power for later measures means that these measures will show larger Type M errors creating the illusion that late effects are bigger than they really are.  Third, late measures are more likely to show Type S errors than early measures, creating the illusion that the direction of effects is more variable in late measures than in early measures.

Taken together, these issues show that power is an important factor in the interpretation of the results of eye-tracking experiments -- even when significant effects were found -- and that researchers should always aim to conduct high power studies to avoid not just Type II errors but also Type S and Type M errors and the three illusions discussed above.  Boosting effect size is the most powerful way to improve power but usually only possible within narrow limits.  The more viable alternative is to increase the number of tested subjects and items.

\subsection{The role of the tested hypotheses}

We demonstrated that the Bonferroni correction results in a smaller loss of power than is commonly believed and that power can be restored by moderately increasing the number of subjects and items.  However, in the long run, the cost of collecting more data adds up and this raises the question of whether the issue can be avoided altogether.  Fortunately, the answer is yes.  The root of the problem is that we often do not have precise enough hypotheses about how a manipulation affects eye movements and, for this reason, we test multiple variants of a hypothesis (effect in measure 1, effect in measure 2, etc.).  If we had more precise predictions, we could cut down on the number of tests and the multiple comparison problem would go away or at least be alleviated.  For example, if a researcher predicts that the effect of a manipulation should emerge in a late measures rather than an early measure, it would be sufficient to only test go-past time and total reading time.  The correction then only needs to account for two comparisons, which preserves more power than a correction for tests in all standard measures.

Research focusing on eye movement control in reading is doing a fairly good job at formulating precise hypotheses.  Computational models of reading such as E-Z Reader and SWIFT make precise predictions not just about the measure in which an effect is expected to appear but also about the size of the effects \cite{ReichleEtAl1998, EngbertEtAl2005}.  In contrast to that, psycholinguistic research is having greater difficulties formulating precise predictions at the level of granularity achieved by models like E-Z Reader and SWIFT.  This partly lies in the nature of the studied processes, which are not as tightly linked to eye movement control, but perhaps there is also some reluctance to engage with the nitty-gritty details of reading behaviour.

A recent example showing that precise linguistic predictions are possible is the work by \citeNP{EngelmannEtAl2013}, who presented a computational model that links sentence processing to the mechanics of eye movement control.  This enables the model to produce the precise predictions that allow us avoid multiple testing \cite<see>[for further discussion]{VasishthEtAl2013}.

\subsection{The role of replication}
\label{the-role-of-replication}

One objection to our conclusions has been that replication is a better method for establishing whether an effect is true or not than applying corrections for multiple comparisons.  We agree that the replication of initial positive findings is important.  \citeNP{McElreathSmaldino2015} show that the veracity of many true hypotheses can only be determined through several iterations of replications.  However, corrections for multiple comparisons and replication are not mutually exclusive.  In fact, the opposite is the case.  A researcher considering to attempt the replication of a result should not base this decision on an uncorrected analysis because the results of that analysis are misleading.  Instead they should base this decision on the corrected results which are much more informative about the expected utility of a replication.

% TODO Cite CramerEtAl2016 and add discussion on what happens with multiple regions of interest and more complex designs.

\subsection{Conclusions}
\label{conclusions}

Thanks to the efforts of the Open Science Collaboration (\citeyear{OpenScienceCollaboration2015}), we know that about 60\% of published results in psychology do not replicate.  The field has not yet found an effective response to this problem but a number of efforts are under way.  For instance, the small but vocal open science movement wants to increase replicability through transparency \cite{ArsendorpfEtAl2013}.  However, how practical and effective these efforts will be remains to be seen.

In this paper, we make a concrete proposal about how the replicability of reading experiments can be improved.  We found that the established standard practice of testing multiple dependent eye-tracking measures without correcting for multiple comparisons raises the rate of false positive results to unacceptable levels even when only four measures are tested.  Studies that test considerably more measures are not uncommon and will be even more affected by this problem.

Fortunately, our results also show that counter-measures like the Bonferroni correction are effective at keeping false positives in check and thus considerably improve the replicability of published results.  These corrections come at the price of reduced statistical power but our simulations show that the loss in power is by far not as dramatic as is often claimed.  A moderate increase in the number of tested subjects and items can make up for the lost power.

Beyond that our results highlight the important but underappreciated role of statistical power.  Specifically, we showed that a failure to detect true effects is not the only risk associated with insufficient statistical power.  Even if effects are detected, low power can give rise to statistical illusions that can lead a researcher to draw incorrect conclusions.

\bibliography{manuscript}

\end{document}